# High-Fidelity Control of Superconducting Qubits Using Direct Microwave Synthesis in Higher Nyquist Zones

**WILLIAM D. KALFUS (Member, IEEE), DIANA F. LEE (Member, IEEE), GUILHEM J. RIBEILL, SPENCER D. FALLEK, ANDREW WAGNER, BRIAN DONOVAN (Member, IEEE), DIEGO RISTÈ, AND THOMAS A. OHKI**

Quantum Engineering and Computing Group, Raytheon BBN Technologies, Cambridge, MA 02138 USA

Corresponding authors: William D. Kalfus (william.d.kalfus@raytheon.com) and Thomas A. Ohki (thomas.ohki@raytheon.com)

This work was supported in part by the Office of the Director of National Intelligence, Intelligence Advanced Research Projects Activity, through the Army Research Office under Contract W911NF-14-1-0114 and in part by internal research support from Raytheon BBN Technologies.

**ABSTRACT** Control electronics for superconducting quantum processors have strict requirements for accurate command of the sensitive quantum states of their qubits. Hinging on the purity of ultra-phase-stable oscillators to upconvert very-low-noise baseband pulses, conventional control systems can become prohibitively complex and expensive when scaling to larger quantum devices, especially as high sampling rates become desirable for fine-grained pulse shaping. Few-gigahertz radio-frequency (RF) digital-to-analog converters (DACs) present a more economical avenue for high-fidelity control while simultaneously providing greater command over the spectrum of the synthesized signal. Modern RF DACs with extra-wide bandwidths are able to directly synthesize tones above their sampling rates, thereby keeping the system clock rate at a level compatible with modern digital logic systems while still being able to generate high-frequency pulses with arbitrary profiles. We have incorporated custom superconducting qubit control logic into off-the-shelf hardware capable of low-noise pulse synthesis up to 7.5 GHz using an RF DAC clocked at 5 GHz. Our approach enables highly linear and stable microwave synthesis over a wide bandwidth, giving rise to high-resolution control and a reduced number of required signal sources per qubit. We characterize the performance of the hardware using a five-transmon superconducting device and demonstrate consistently reduced two-qubit gate error (as low as 1.8%), which we show results from superior control chain linearity compared to traditional configurations. The exceptional flexibility and stability further establish a foundation for scalable quantum control beyond intermediate-scale devices.

## I. INTRODUCTION

Quantum computing is widely regarded as a promising technology for solving classically intractable problems in fields ranging from chemistry [1] to cryptography [2]. Superconducting qubits have emerged as a viable candidate for a physical realization of a quantum computer [3], [4], as they are able to be fabricated using well-developed techniques from the semiconductor industry and are controlled using standard microwave electronics. However, state-of-the-art qubits are quite "noisy" compared to the error rates required for quantum algorithms to be carried out with high fidelity [5]; this can be mitigated by combining many noisy qubits into fewer reduced-noise logical qubits through quantum error correction [6], [7]. Additionally, certain algorithms combat noise by directly utilizing higher numbers of qubits, such as variational quantum optimizers [8]. These techniques can rapidly increase the number of qubits that a quantum algorithm requires, and the electronics that enable these interactions grow increasingly complex and expensive [9].

The microwave electronics that comprise a control system for superconducting qubits must have extremely low electrical noise so as not to reduce coherence [10]. This requirement is even more critical when driving all-microwave two-qubit gates [11]–[13], in which one instrument channel may need to remain phase coherent with the rotating reference frames of multiple qubits, making tight phase locking between channels essential. With conventional control hardware requiring as many as four arbitrary waveform generator channels per qubit (two dual-quadrature complex pairs) [4], distributing a stable clock reference, suppressing spurious tones, and adding infrastructure to calibrate imperfections (such as local oscillator (LO) and sideband leakage) in tens or hundreds of instrument channels becomes a daunting task as the density of quantum hardware increases.

It is often difficult to predict how much power must be synthesized at room temperature to drive any individual qubit before characterizing it, even before attempting to reduce gate time to optimize higher depth quantum algorithms. Shorter gates require higher amplitude [14], underscoring the need for a control system to be capable of pulse synthesis over a wide dynamic range with minimal distortion. Furthermore, all-microwave two-qubit gates can require as much as ten times the amplitude of single-qubit gates (see Section IV). Although modern mixers used in upconversion systems can behave approximately linearly over a moderate voltage range [15], the true output diverges from this approximation at high input power (determined by the manufacturer). Avoiding the resulting harmonics is often achieved by either using mixers with higher power ratings or by using lower input powers and increasing amplification at the output; however, the former solution often requires specialized components with limited bandwidths, and the latter solution requires additional low-noise amplifiers, which also distort high-power signals. Additionally, attenuating the input signal degrades the signal-to-noise ratio at the output (as well as the usable vertical resolution of the pulse synthesizer if passive attenuation is not used) and other mixer nonidealities, such as LO leakage and imperfect sideband suppression are amplified [16]. Furthermore, the electrical properties of mixers tend to vary due to manufacturing tolerances and environmental effects, and therefore periodic calibration must be performed to sufficiently suppress the microwave carrier and unwanted image tones [17]. Performing these calibrations requires additional hardware to probe the desired outputs and places additional overhead on long-running experiments.

The commercial availability of radio-frequency digital-to-analog converters (RF DACs) with multi-gigahertz sampling rates [18], [19] presents a promising alternative to traditional microwave pulse synthesis for superconducting qubits in terms of channel count, spurious emissions, noise performance, and distortion. Whereas signal generators that directly approximate the microwave pulse are capable of controlling a superconducting qubit [20], [21], doing so with acceptable resolution at typical qubit and resonator frequencies (approximately 5–10 GHz) requires sampling at multiple tens of gigahertz, hindering scaling to larger devices. However, an off-the-shelf wide-bandwidth DAC with a sampling rate slightly below desired frequencies can be used to synthesize pulses in higher Nyquist zones, thus maintaining the scalability of a direct synthesis solution without significant overhead compared to conventional systems.

Having a wide-bandwidth control instrument can also greatly simplify the process of implementing modern techniques for improving fidelity in quantum circuits, such as optimal control [22]–[24], parametric gates [25]–[27], and entanglement stabilization protocols [28]. Additionally, certain quantum error correction codes require driving multiple modes of an oscillator [29], [30], which could be performed using a single wide-bandwidth channel in lieu of combining multiple channels, thereby reducing the implementation cost of these schemes.

In this article, we demonstrate the capability of an RF DAC operating in higher Nyquist zones to synthesize shorter, lower error, all-microwave two-qubit gates than those generated by a state-of-the-art upconversion system. We demonstrate control and readout of a qubit using a single instrument channel and show superior linearity over a wide dynamic range compared to the upconversion system.

## II. THEORY

Modern innovation in wireless communications has given rise to a market for affordable wide-bandwidth DACs, making it possible to integrate an off-the-shelf component and exploit the fundamental presence of tones above its clock rate. To elucidate the origin of these tones, we consider an analytical treatment of the DAC as a device which accepts numerical data ("samples") through a digital interface and converts them into a corresponding voltage. It is assumed that there is a linear relationship between the numerical value of a sample and the corresponding analog output, and that samples are converted at a constant rate $f_s$ [31]. When samples are generated by computing the value of an analytical function $x(t)$, the output of the DAC may be expressed as

$$v(t) = \left[ x(t) \sum_{k=-\infty}^{\infty} \delta(t - kT) \right] * r(t) \quad (1)$$

where $T = 1/f_s$ is the sampling period and $(*)$ represents convolution in time. The "reconstruction waveform" $r(t)$ describes how the output of the DAC behaves between samples and is nonzero only for $0 \leq t < T$. The corresponding frequency spectrum of the DAC output is

$$V(\omega) = R(\omega) \left[ X(\omega) * \sum_{n=-\infty}^{\infty} \delta(\omega T - 2\pi n) \right] \quad (2)$$

where capital-letter functions denote the Fourier transform (FT) of the corresponding time-domain function (a derivation of this expression is provided in Appendix A). The summation represents a series of peaks at all integer multiples of the sampling frequency, and the convolution of $X(\omega)$ with these peaks creates copies of the spectrum of the original signal at each of these peaks [see Fig. 1(a)]. If $X(\omega)$ has a bandwidth greater than $f_s/2$, these copies will overlap, resulting in aliasing. Consecutive bands of width $f_s/2$ are referred to as Nyquist zones, with the first Nyquist zone spanning DC to $f_s/2$, the second Nyquist zone spanning $f_s/2$ to $f_s$, etc. Hence, when a signal is confined to a single Nyquist zone, sampling the signal places a copy of the signal's spectrum in each odd Nyquist zone and an inverted copy in the even Nyquist zones [31].

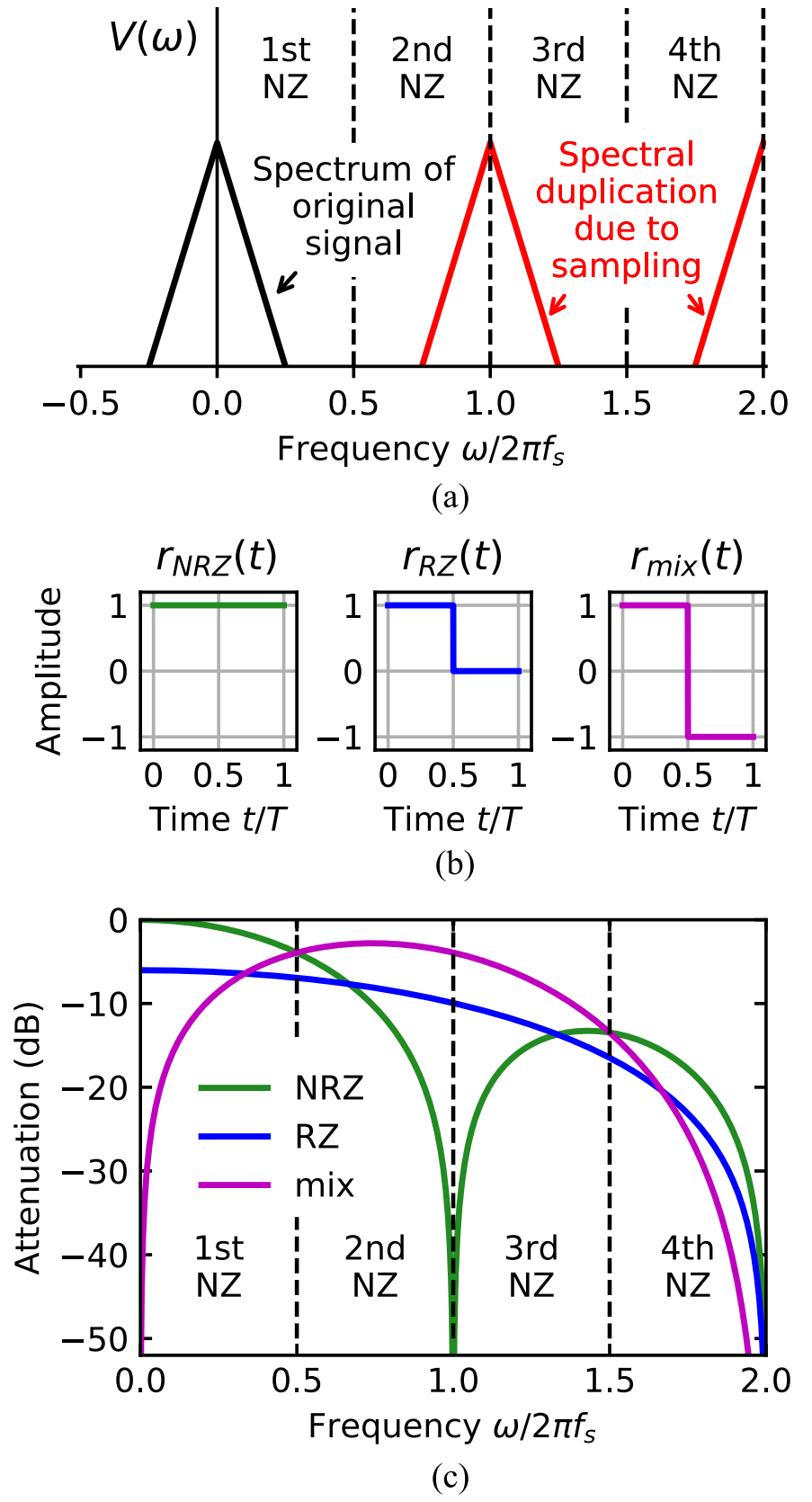

**FIGURE 1. (a) Arbitrary signal spectrum at baseband and duplicated spectra resulting from sampling the original signal at sampling rate $f_s$. The signal bandwidth is confined to the first Nyquist zone (NZ), or less than $f_s/2$, to prevent aliasing during sampling. (b) Shapes of three common reconstruction waveforms available in RF DACs: non-return-to-zero (NRZ), return-to-zero (RZ), and mix/RF-mode. (c) Frequency-dependent attenuation due to each reconstruction waveform. Different reconstruction waveforms may be chosen to optimize the system for higher power in desired Nyquist zones (see Appendix A).**

The reconstruction waveform $r(t)$ manifests itself as a frequency-dependent attenuation, but careful choice of $r(t)$ can be used to maximize power in preferred Nyquist zones. Three common functions for $r(t)$ include non-return-to-zero (NRZ), return-to-zero (RZ), and mix-mode/RF-mode [see Fig. 1(b)], each with their own attenuation profile [see Fig. 1(c)]. Analytical expressions and their derivations are provided in Appendix A. The NRZ waveform is the most common and is often the only choice available in low-frequency DACs. Also referred to as a zero-order hold, this reconstruction waveform represents the DAC presenting a sample at its output and holding that value constant until the next sample [31]. More advanced DACs can update their outputs between samples, such as driving the output to zero or inverting the output. When this occurs exactly halfway between samples, the former describes RZ mode and the latter describes mix mode. Although this requires the analog core of the DAC to have a higher sampling rate, mathematically this is treated as a series of impulses at the original sampling rate with nonconstant behavior between samples. Because our hardware uses DACs with sampling rates near qubit and resonator frequencies, the second and third Nyquist zones are the desired bands for signal synthesis, and choosing mix mode yields the highest power in this range. We note that other reconstruction waveforms could be employed to further optimize power output in various Nyquist zones, but we choose not to discuss these as they are less common.

## III. HARDWARE

We implement direct microwave synthesis using a combination of off-the-shelf hardware and custom field-programmable gate array (FPGA) logic [see Fig. 2(a)]. We employ the VadaTech AMC599 [32], which contains a Xilinx XCKU115 UltraScale FPGA [33] and two Analog Devices AD9164 DACs [34] clocked at 5 GHz in an advanced mezzanine card (AMC) form factor module. We note that the actual data stream of the DAC is limited to 2.5 GSa/s due to the use of complex data, but because the update rate of the DAC is still 5 GHz, this does not affect the location of the Nyquist zones. Because the data are complex, tones may be generated directly as upper or lower sidebands of a carrier produced by an NCO internal to the DAC; when this NCO is centered in an Nyquist zone, a signal bandwidth of 2.5 GHz is achieved. Multiple AMC599 modules are loaded into a VadaTech VT848 1U AMC chassis [35].

The FPGA logic commanding the DACs iterates upon our previous work developing specialized instrumentation for superconducting qubit experiments [36]. Unique functionality in the system includes a sequencer for advanced control flow, waveform engines with multiple NCOs for coherent control between densely connected qubits, and marker engines for triggering external instruments. An embedded processor configures clocking, DACs, and external communication interfaces.

To demonstrate the efficiency of the RF DAC, individual control and readout is achieved with a single channel per qubit. This is made possible by the architecture of our superconducting device, in which all qubits are driven through their transmission line readout resonators. With only a single externally coupled feedline per qubit, the RF DAC offers an efficient alternative to conventional upconversion systems requiring two dual-quadrature baseband signals to be upconverted with IQ mixers driven by separate microwave sources. An analog front-end for the RF DAC is constructed as shown in Fig. 2(c) (full wiring shown in Appendix D). The DAC output is split and selectively filtered; this allows for extraction of the readout pulse for heterodyne downconversion and for additional attenuation without sacrificing dynamic range. Multiple amplifiers are employed to provide enough power to apply gates and to drive the LO port of the downconversion mixer with the synthesized readout pulse. The control

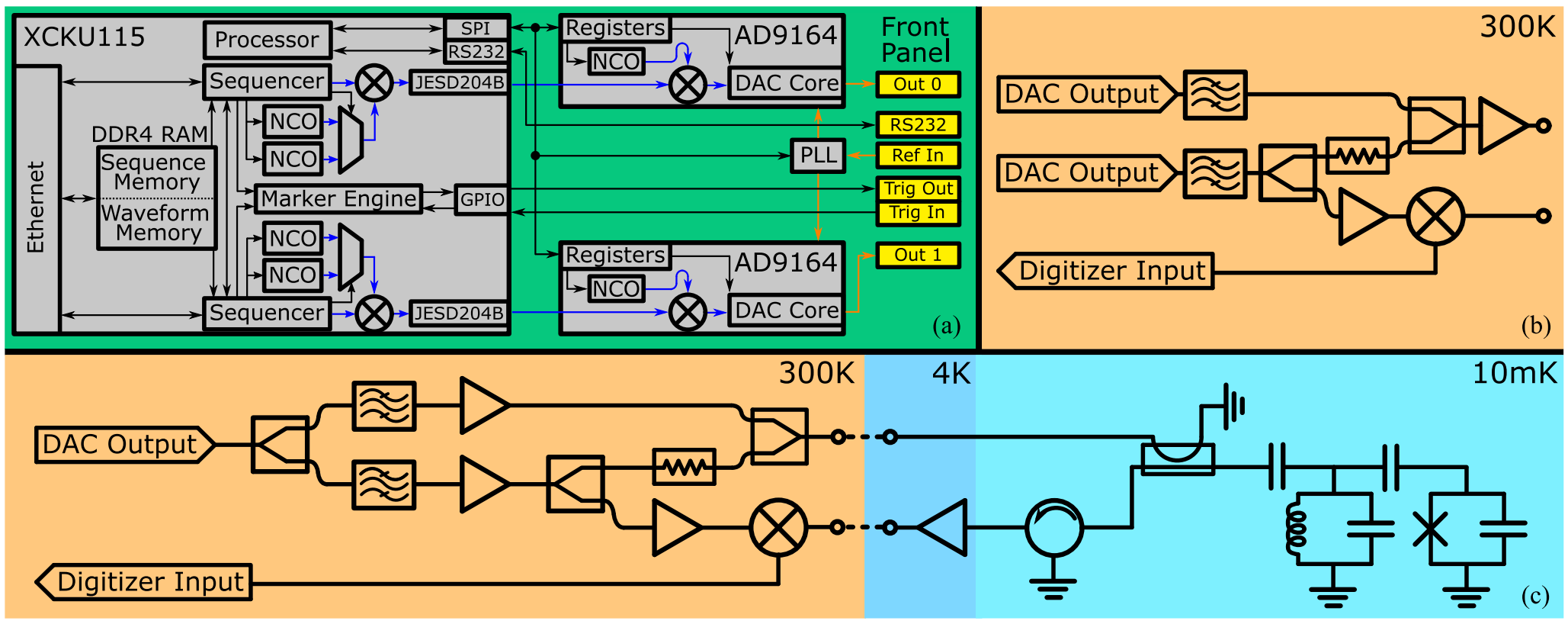

**FIGURE 2. (a) Internal block diagram of direct RF synthesis system. Blue lines indicate paths carrying dual-quadrature digital information, and orange paths indicate analog signals. RS232 is exposed on the front panel for communication with the embedded processor. The processor configures numerically controlled oscillators (NCOs) and reconstruction waveforms in the RF DACs by writing to digital registers using a serial peripheral interface (SPI). The onboard phase-locked loop (PLL) exposes an input for a reference clock from which the RF DAC and FPGA clocks are derived. The FPGA logic implements two sequencers that read instructions and waveforms from memory, numerically modulate baseband tones, and stream the complex data to the RF DACs over JESD204B. Instructions and waveforms are written to memory over Ethernet, which is also used to configure the sequencers. (b) Topology of dual-channel control and readout hardware used to synthesize two-qubit gates (complete wiring used for experiments available in Appendix C). (c) Topology for single-channel control and readout (complete wiring used for experiments available in Appendix D).**

and readout paths are then recombined before entering the cryogenic environment.

## IV. RESULTS

We compare the performance of the RF DAC to a traditional control scheme by conducting experiments on a device with five fixed-frequency transmon qubits [7] located at the base of a dilution refrigerator operating at 10 mK. Our upconversion control system uses the BBN second-generation Arbitrary Pulse Sequencer (APS2) [36] to generate dual-quadrature signals at 1.2 GSa/s. Control pulses are then upconverted with a single channel of a Holzworth HS9003A microwave source shared between all qubits, whereas measurement pulses are upconverted using separate sources (Vaunix Lab Bricks) for each qubit. All instruments are phase-locked using a Stanford Research Systems FS725 Rubidium Frequency Standard. Device details and a complete diagram of the upconversion system are provided in Appendix C.

We first calibrate the $X_{\pi/2}$ rotation gate comprised of a Gaussian pulse with DRAG [37], [38] for one qubit on the device using both control systems. Because the RF DAC and upconversion system operate at different clock rates, pulse lengths must differ by a small amount; single-qubit gates are calibrated at 50 ns when using upconversion and at 48 ns when using direct RF synthesis.

To confirm that the RF DAC does not degrade relevant qubit performance metrics, we measure the characteristic decay time $T_{2e}$ of a spin echo experiment and single-qubit error per Clifford (as measured by randomized benchmarking [39]) and find comparable values to those obtained when using upconversion (see Fig. 3). In particular, the lack of any additional dephasing indicates that the integrated phase noise of the RF DAC over the coherence time of the qubit is comparable to that of an upconversion system that utilizes an ultra-low-phase-noise local oscillator [10]. We have observed that the energy relaxation time $T_1$ is also undiminished when using the RF DAC (approximately 60 $\mu$s; see Appendix C for measurements taken during two-qubit experiments). Hence, the energy relaxation limit for single-qubit gate error is approximately $4 \times 10^{-4}$, an order of magnitude below the single-qubit error we measure.

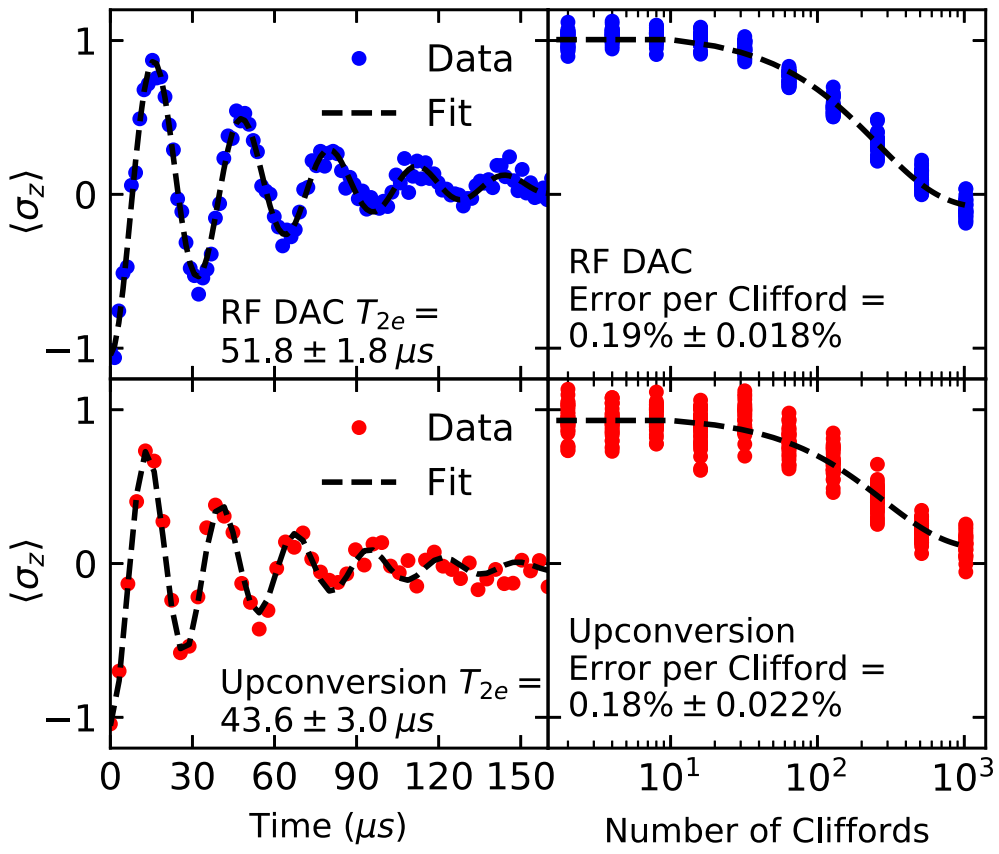

**FIGURE 3. Spin echo (left column) and single-qubit randomized benchmarking (RB) (right column) using direct RF synthesis (top row) and upconversion (bottom row). Phase offsets are introduced to the spin echo to improve the quality of the decay fit. Performance is not degraded when using the RF DAC compared to that of upconversion; discrepancies between exact values are likely due to natural time-varying fluctuations in coherence (see Appendix C).**

Next, we compare the performance and stability of the RF DAC when driving two-qubit gates to that of the upconversion system. Our device uses all-microwave two-qubit echoed cross-resonance (CR) gates between qubits directly coupled through a bus resonator [11], [40], [41]. The gate is comprised of two flat-top Gaussian pulses driven at the target qubit frequency through the control channel, along with an $X_\pi$ rotation on the control qubit to cancel undesired terms the CR Hamiltonian. We compute two-qubit gate error as the mean of the target and control error measured using two-qubit RB [42], and we choose to use the qubit pair with the lowest two-qubit error rate on the device.

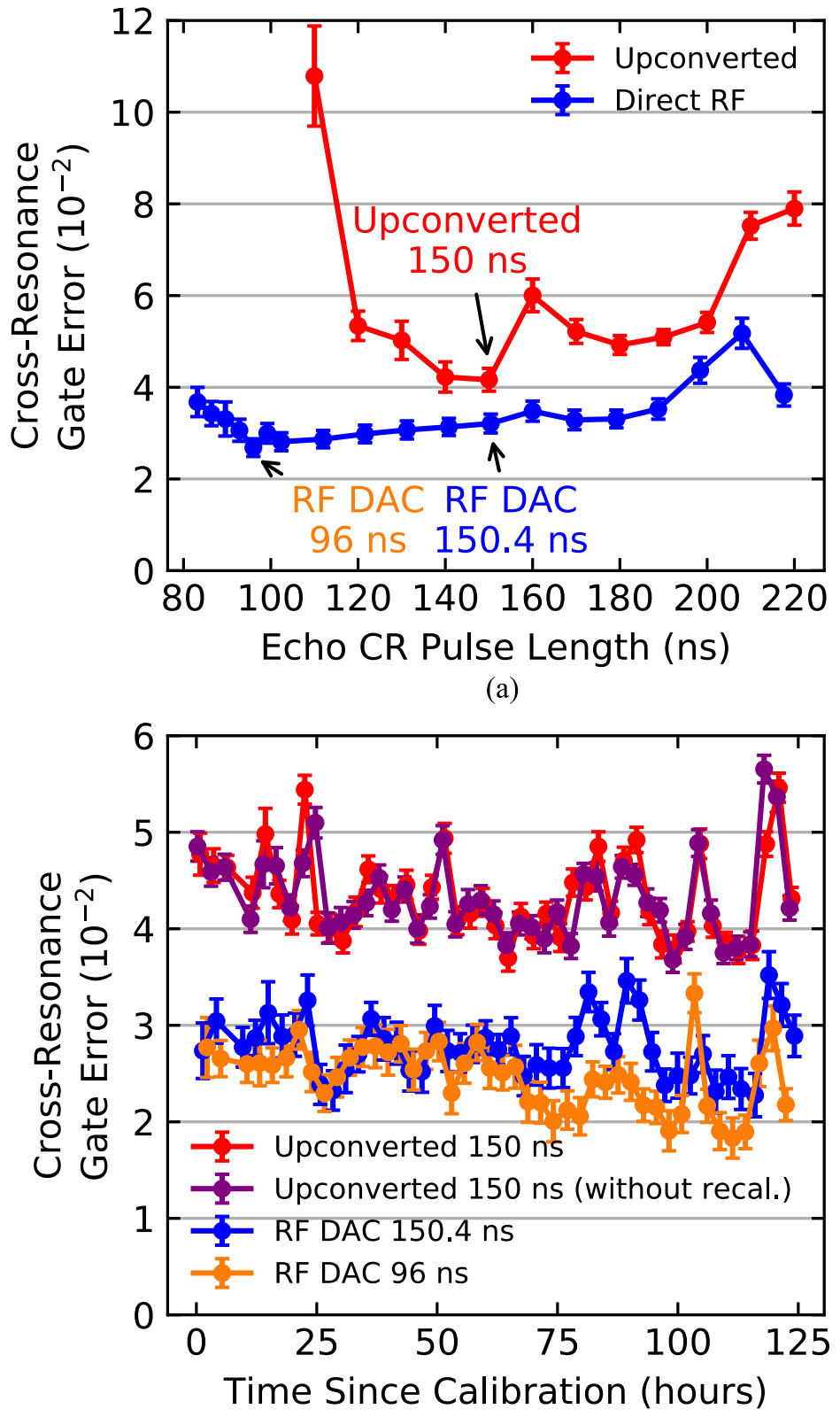

(b)

**FIGURE 4. (a) CR gate error as a function of pulse length, computed as the mean error between target and control qubits. Pulses are flat-topped Gaussians with $2\sigma$ rise/fall time, with $2\sigma = 19.2$ (20) ns for the RF DAC (upconversion system). Error rates are averaged across 8–11 independent calibrations, and error bars are computed as the root mean square of the fit error across all gate error measurements. (b) Stability of error rates over time for a 150-ns upconverted gate (with and without calibrating mixers before each instance of RB, corresponding to red and purple curves, respectively), a 150.4-ns gate synthesized with the RF DAC (blue curve), and a 96-ns gate synthesized with the RF DAC (orange curve). Points are computed as the mean error between target and control qubits, and error bars correspond to the exponential decay fit error. These gates correspond to the points denoted in (a).**

To reduce the number of external components required, for experiments with CR gates, we assemble an analog front-end using separate DACs for control and readout as shown in Fig. 2(b) (full wiring is presented in Appendix C). We then sweep the length of the CR pulse, and for each pulse length, the gate is calibrated and 30 sequences of two-qubit RB are used to characterize the quality of the calibration. Between 8 and 11 repetitions of this process are performed at each pulse length to avoid local minima in gate error due to small variations in calibration parameters.

The RF DAC is able to synthesize shorter pulses at higher amplitudes with gate error remaining consistently below 4% [see Fig. 4(a)], whereas the upconverted gate error rapidly exceeds 10% as pulse length decreases. The ability of the RF DAC to synthesize short gates is especially beneficial for high-depth quantum algorithms that depend on executing numerous entangling operations within the qubits' coherence times.

For many types of two-qubit gates, if the relative phases between instrument channels drift, those channels will require recalibration in order to rotate the target state around the proper axis [11]. To characterize the extent to which naturally occurring phase drift impacts gate performance, we calibrate CR gates of multiple lengths on both control systems and track their error rates over time by repeatedly measuring identical random sequences without pulse recalibration. To decouple the effect of inherent time-varying fluctuations, we use mechanical microwave switches to interleave experiments with each control system. The upconversion mixers are calibrated once at the start of the experiment, and error is measured over time both using the initial calibration parameters and after recalibrating them before each instance of RB. A 150-ns gate was chosen for the upconversion system as it showed minimal error and close performance to the RF DAC at a similar pulse length [see Fig. 4(a)]. A 150.4-ns gate and a 96-ns gate are tracked for the RF DAC in order to examine the performance of both the most similar gate to the one chosen for the upconversion system as well as the one that exhibited the lowest error.

Gates synthesized by the RF DAC exhibit a consistent reduction in two-qubit gate error that is maintained over the course of 125 hours [see Fig. 4(b)], reaching a minimum of 1.8% (compared to 3.8% for the recalibrated upconversion system at a similar point in time). Significant fluctuations are consistent between the two control systems, indicating that such events are not due to particular deleterious effects of either system. Importantly, the results of this experiment indicate that separate RF DACs can maintain phase coherence with each other over long timescales without degrading two-qubit gate error compared to sharing a single local oscillator among upconversion channels.

To investigate the divergence in error between the RF DAC and the upconversion system at short pulse lengths, we characterize the error rates of CR gates at various power levels and compare it to the linearity of the two control systems at similar power levels. We consider linearity instead of other metrics (such as phase noise) as we do not observe significant differences in coherence or single-qubit gate fidelity [10], [43]. We quantify the linearity $L$ of a control system as $L(A_n) = \frac{dV_o}{dA}|_{A=A_n}$, where $V_o$ is the output voltage and $A$ is the programmed amplitude of the synthesizer, which we measure for various amplitudes $A_n$; in a perfectly linear system, $L$ would be a constant. We measure linearity for the RF DAC and for the output of the upconversion mixer, as well as for a single quadrature in the upconversion setup in order to decouple any nonlinearity arising from baseband amplification. Additional metrics [such as spurious-free dynamic range (SFDR)] are discussed in Appendix B.

As linearity decreases, two-qubit gate error starts to rapidly increase [see Fig. 5(a)], providing evidence for the strong sensitivity of gate performance to control system linearity. Comparing the results between the mixer output and a single input quadrature shows that the performance degradation is due to the mixer rather than any prior amplification; hence, even with extremely high-performance pulse synthesizers and local oscillators, the presence of a single nonlinear circuit element can degrade performance considerably.

To demonstrate the deleterious effects of nonlinearity in a control system, we measure the output voltage spectra at high amplitude. Harmonics have a strong presence in the output of the upconversion system, and because they have the same spectral profile as the primary pulse, they are not confined to a negligible bandwidth [see Fig. 5(c)]. The resulting low-amplitude tails of harmonics in the CR pulse overlap with the frequency space of the control qubit, and the harmonics of the $X_{\pi/2}$ pulse overlap with the frequency space of the target qubit. Though they are suppressed at low amplitude, the harmonics never fully vanish due to the inherent nonlinearity of the circuit [16]. This is especially detrimental for devices that are sensitive to single-photon perturbations, such as readout resonators [4], and attempting to achieve an optimal detuning for a CR gate [44] while avoiding a series of harmonics as well as LO leakage becomes a difficult task, especially for devices with higher connectivity.

Conversely, the RF DAC, being a linear control element, shows no spectral components except at desired locations [see Fig. 5(d)]. This further emphasizes the usefulness of the RF DAC for experiments that require spectral purity and simplifies the considerations needed when scaling to larger quantum devices.

## V. CONCLUSION

We have demonstrated that an RF DAC operating in higher Nyquist zones can control qubits with fidelities exceeding those of a typical upconversion-based control system as well as perform measurement without requiring an additional instrument channel, creating a scalable platform that avoids fundamental imperfections and limitations of nonlinear hardware. Hence, the RF DAC can drive shorter gates than the upconversion setup while maintaining low error due to the increased linearity of the system, avoiding the need for an additional channel dedicated to high-amplitude pulses. The gate error of the RF DAC is stable over long timescales, indicating good phase stability, and the lack of harmonic content in its output greatly simplifies frequency planning. These observations emphasize the utility of an RF DAC in creating a flexible, scalable control system for superconducting qubits beyond the intermediate scale.

The clean spectrum of the RF DAC offers a promising solution to frequency crowding, which is a major barrier to scaling quantum devices to many qubits; preventing collisions of the upconversion products, LO leakage, and imperfect sideband suppression with all qubit transitions—often in a tight frequency range to optimize CR gate speed—becomes increasingly difficult for large devices [45]–[47]. Furthermore, even if such careful frequency planning were carried out, gate error is often not a flat function of sideband frequency, preventing optimal performance from being achieved solely due to the use of an upconversion system.

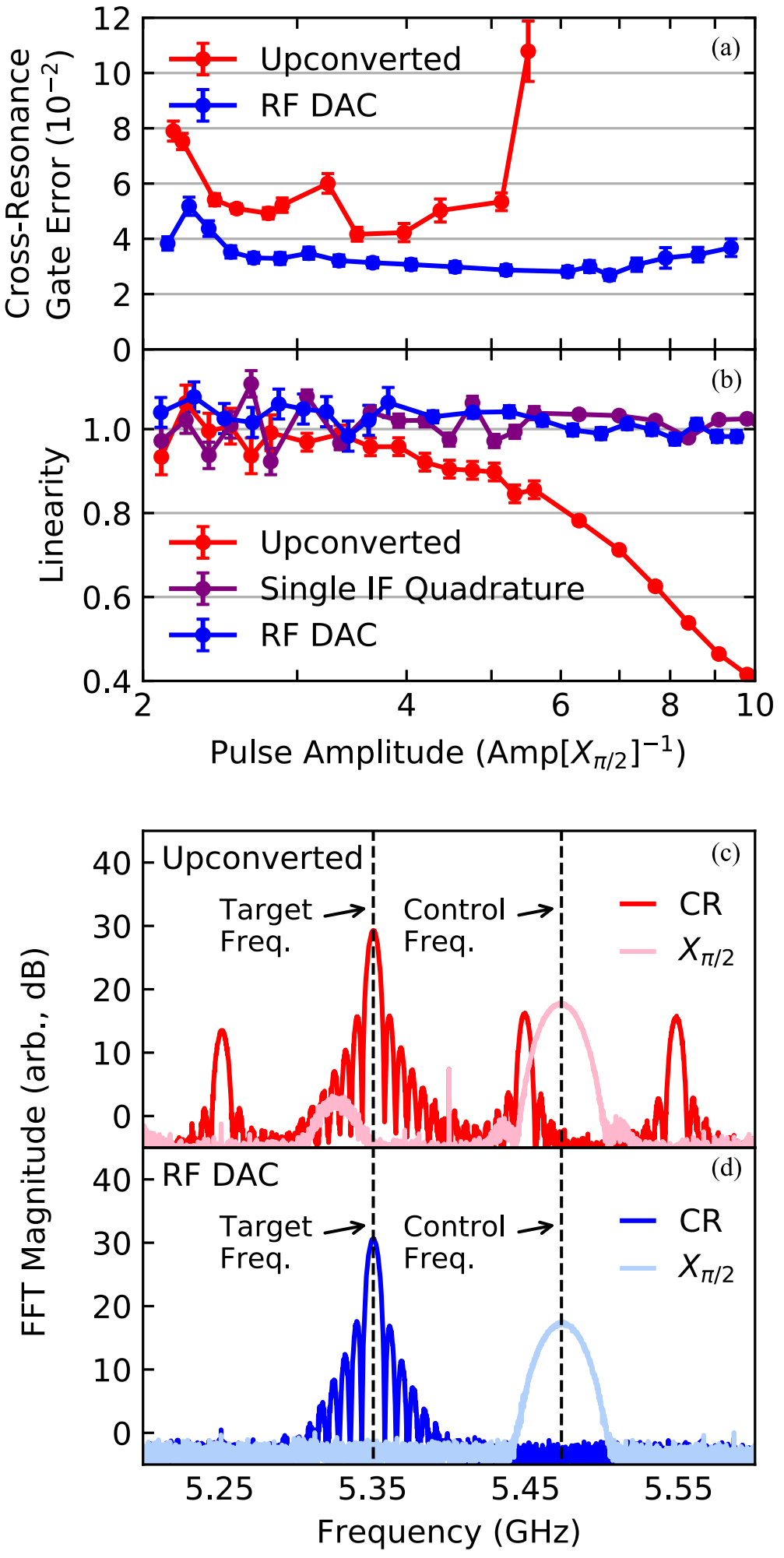

**FIGURE 5. (a) CR gate error as a function of pulse amplitude. Amplitudes are normalized against the amplitude of the control qubit $X_{\pi/2}$ gate for each control system. (b) Control system linearity at various pulse amplitudes. Linearity is normalized against its average value at low amplitudes (between 10%–20% of full-scale system amplitude). Comparing to (a), gate error increases as linearity decreases. (c) Spectra of control qubit drive output for upconverted CR and $\pi/2$ pulses. Distortion induced by nonlinear circuit elements creates harmonics in the output, and overlap between critical frequency bands can be observed. (d) Spectra of control qubit drive output for CR and $\pi/2$ pulses synthesized with the RF DAC. The spectrum shows no spurious emissions or other undesired artifacts. In both (c) and (d), the frequencies of the target and control qubits are both denoted by dashed lines.**

We believe that the consistent improvement in CR gate fidelity demonstrated by the RF DAC results from the high amplitude required compared to single-qubit gates; any spurious emissions in the signal will have a stronger impact on the driven qubits and on surrounding qubits (due to unintentional coupling and classical crosstalk) even when significantly suppressed, only to be further exacerbated by increasing power to reduce gate length. Since spurious emissions are always present in an upconversion system even at low power due to the fundamental nonlinear nature of mixers, and since no particular spurious emission can be designated as the sole cause of the error (especially due to the nonnegligible rolloff of the spectral profile taken on by the spurious tones), using an RF DAC with a clean spectrum avoids this undesired behavior entirely at all power levels, leading to improved gate fidelity.

While in this article, we have demonstrated the capability of off-the-shelf hardware, certain engineering challenges remain in order to take full advantage of the flexibility offered by the RF DAC. In particular, the output of the DAC must be filtered to prevent interference from tones in Nyquist zones other than the one(s) desired and from spurious emissions at digital clock frequencies present in the system; in this work, we mitigate these effects using commonly available microwave filters chosen to match the frequency bands of interest. Future work will address reconfigurable filtering electronics for flexible integration as well as dynamic adjustment of digital clocks, allowing for optimal placement of Nyquist zones (many system architectures utilize digital PLLs for generating clock signals, enabling software-based adjustment).

## APPENDIX A
## FREQUENCY SPECTRUM OF SAMPLED SIGNALS AND RECONSTRUCTION WAVEFORMS

We derive the synthesis of tones in higher Nyquist zones by first considering the voltage at the output of a DAC when playing samples of an analytical function $x(t)$ presented in (1)

$$v(t) = \left[ x(t) \sum_{k=-\infty}^{\infty} \delta(t-kT) \right] * r(t)$$

where $T$ is the sampling period, $r(t)$ is the reconstruction waveform, and $(*)$ denotes convolution in time. We start computing the FT of this expression by applying multiplication and convolution identities

$$\begin{aligned} V(\omega) &= \mathscr{F}\{v(t)\} \\ &= \left[ X(\omega) * \mathscr{F}\left\{ \sum_{k=-\infty}^{\infty} \delta(t-kT) \right\} \right] R(\omega). \end{aligned} \quad (3)$$

In this expression, $X(\omega) = \mathscr{F}\{x(t)\}$ is the spectrum of the original signal and $R(\omega) = \mathscr{F}\{r(t)\}$ is the spectrum of the chosen reconstruction waveform (derived below). To compute the FT of the impulse train (represented by the summation), we recognize its periodicity in time with period $T$ and express it as a Fourier series

$$\sum_{k=-\infty}^{\infty} \delta(t-kT) \equiv \sum_{n=-\infty}^{\infty} c_n e^{2\pi i n t/T}$$

where $c_n$ are the Fourier series coefficients. We solve for $c_n$ by integrating over one period of the impulse train centered at $t=0$; this reduces the summation to only a single term with $k=0$

$$c_n = \frac{1}{T}\int_{-T/2}^{T/2} \delta(t) e^{-2\pi i n t/T} dt = \frac{1}{T}.$$

Therefore

$$\sum_{k=-\infty}^{\infty} \delta(t-kT) = \sum_{n=-\infty}^{\infty} \frac{1}{T} e^{2\pi i n t/T}.$$

We then solve for the FT of this expression

$$\begin{aligned} \mathscr{F}\left\{ \sum_{n=-\infty}^{\infty} \frac{1}{T} e^{2\pi i n t/T} \right\} &= \frac{1}{T} \sum_{n=-\infty}^{\infty} \mathscr{F}\{e^{2\pi i n t/T}\} \\ &= \frac{1}{T} \sum_{n=-\infty}^{\infty} \delta(\omega - 2\pi n/T) \\ &= \sum_{n=-\infty}^{\infty} \delta(\omega T - 2\pi n). \end{aligned} \quad (4)$$

Finally, we substitute this back into (3) to recover (2)

$$V(\omega) = R(\omega)\left[ X(\omega) * \sum_{n=-\infty}^{\infty} \delta(\omega T - 2\pi n) \right].$$

We can derive $R(\omega)$ for the three aforementioned reconstruction waveforms by directly computing their FT. We start with the NRZ reconstruction waveform

$$r_{\text{NRZ}}(t) = u(t) - u(t-T) \quad (5)$$

where $u(t)$ is the step function. Using the known FT of the step function and the time shifting property of the FT, we can directly compute the FT of $r_{\text{NRZ}}$

$$\begin{aligned} R_{\text{NRZ}}(\omega) &= \mathscr{F}\{r_{\text{NRZ}}(t)\} \\ &= \mathscr{F}\{u(t)\} - \mathscr{F}\{u(t-T)\} \\ &= (1 - e^{-i\omega T})\mathscr{F}\{u(t)\} \\ &= (1 - e^{-i\omega T})(\frac{1}{i\omega} + \pi\delta(\omega)) \\ &= e^{-i\omega T/2}(e^{i\omega T/2} - e^{-i\omega T/2})\frac{1}{i\omega} \\ &= 2ie^{-i\omega T/2} \sin\left(\frac{\omega T}{2}\right) \frac{T/2}{i\omega T/2} \\ &= Te^{-i\omega T/2} \operatorname{sinc}\left(\frac{\omega T}{2}\right) \end{aligned} \quad (6)$$

where $\operatorname{sinc}(x) = \frac{\sin(x)}{x}$.

The RZ reconstruction waveform is given by

$$r_{\text{RZ}}(t) = u(t) - u(t - T/2)$$

this expression is identical to $r_{\text{NRZ}}$ with $T \to \frac{T}{2}$; hence, $R_{\text{RZ}}$ is given simply by applying the same transformation to $R_{\text{NRZ}}$

$$R_{\text{RZ}}(\omega) = \frac{T}{2} e^{-i\omega T/4} \operatorname{sinc}\left(\frac{\omega T}{4}\right). \quad (7)$$

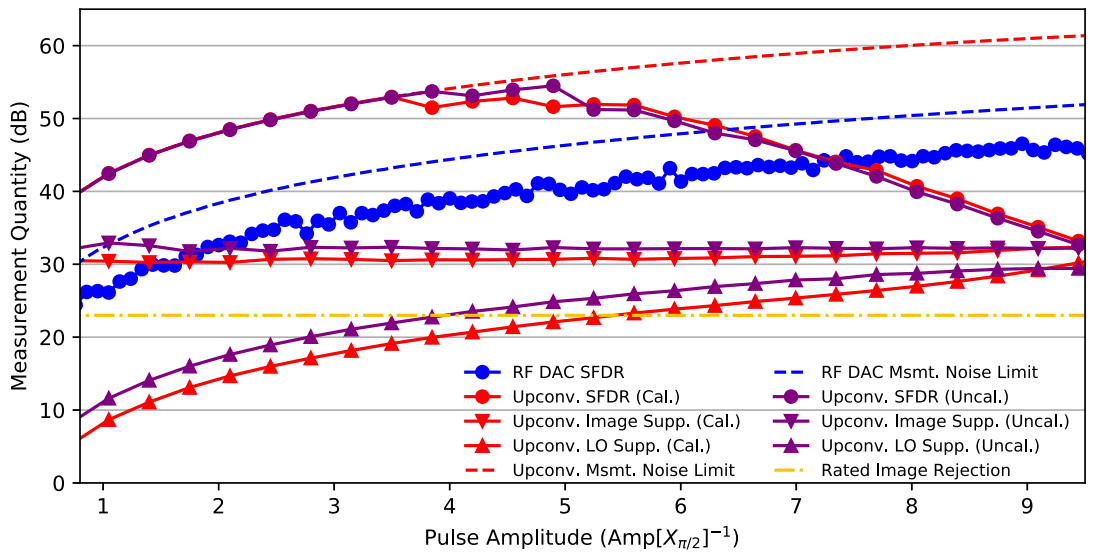

**FIGURE 6. Spectral properties of signals synthesized by the RF DAC (blue) and the upconversion system, both with (red) and without (purple) calibrating the mixer. SFDR (circles) is calculated excluding specific expected emissions for each system, as described in the text. Of those excluded for the upconversion system, the measured (downward-facing triangles) and rated [48] (yellow dash-dot) sideband suppression are presented, as well as the measured suppression of LO leakage compared to the main tone (upward-facing triangles). Dashed lines indicate the noise floor of the SFDR measurements.**

The mix mode reconstruction waveform is given by

$$r_{\text{mix}}(t) = u(t) - u(t - T/2) - [u(t - T/2) - u(t - T)].$$

We start by considering a time shift of $\frac{T}{2}$ on the term in brackets

$$\begin{aligned} R_{\text{mix}}(\omega) &= \mathscr{F}\{r_{\text{mix}}(t)\} \\ &= \mathscr{F}\{u(t) - u(t - T/2)\} \\ &\quad - e^{-i\omega T/2}\mathscr{F}\{u(t) - u(t - T/2)\}. \end{aligned}$$

We recognize the term in the FT as the expression for $r_{\text{RZ}}(t)$ and substitute the transform for $R_{\text{RZ}}(\omega)$

$$\begin{aligned} R_{\text{mix}}(\omega) &= (1 - e^{-i\omega T/2})R_{\text{RZ}}(\omega) \\ &= \frac{T}{2}(1 - e^{-i\omega T/2})e^{-i\omega T/4}\,\text{sinc}\left(\frac{\omega T}{4}\right) \\ &= \frac{T}{2}e^{-i\omega T/2}(e^{i\omega T/4} - e^{-i\omega T/4})\,\text{sinc}\left(\frac{\omega T}{4}\right) \\ &= Tie^{-i\omega T/2}\sin\left(\frac{\omega T}{4}\right)\text{sinc}\left(\frac{\omega T}{4}\right) \\ &= \frac{\omega T^2}{4}e^{-i(\omega T - \pi)/2}\,\text{sinc}^2\left(\frac{\omega T}{4}\right). \end{aligned} \tag{8}$$

The results of (6)–(8) are plotted in Fig. 1(c).

## APPENDIX B
## SPURIOUS EMISSIONS AND LINEARITY

In Fig. 5, we compared CR gate error to pulse distortion resulting from control chain nonlinearity, which we quantified using a linearity metric related to amplitude compression. A common metric for indicating spectral purity of a signal synthesizer is spurious-free dynamic range (SFDR), which may be used to quantify the strength of harmonics arising from nonlinearity. However, this metric alone is not sufficient for inferring a control system's ability to control qubits with low error, as it does not discriminate between linear and nonlinear distortion. To demonstrate this, we measure the SFDR of the upconversion system and the RF DAC as a function of amplitude when synthesizing a continuous wave at the target qubit frequency (see Fig. 6).

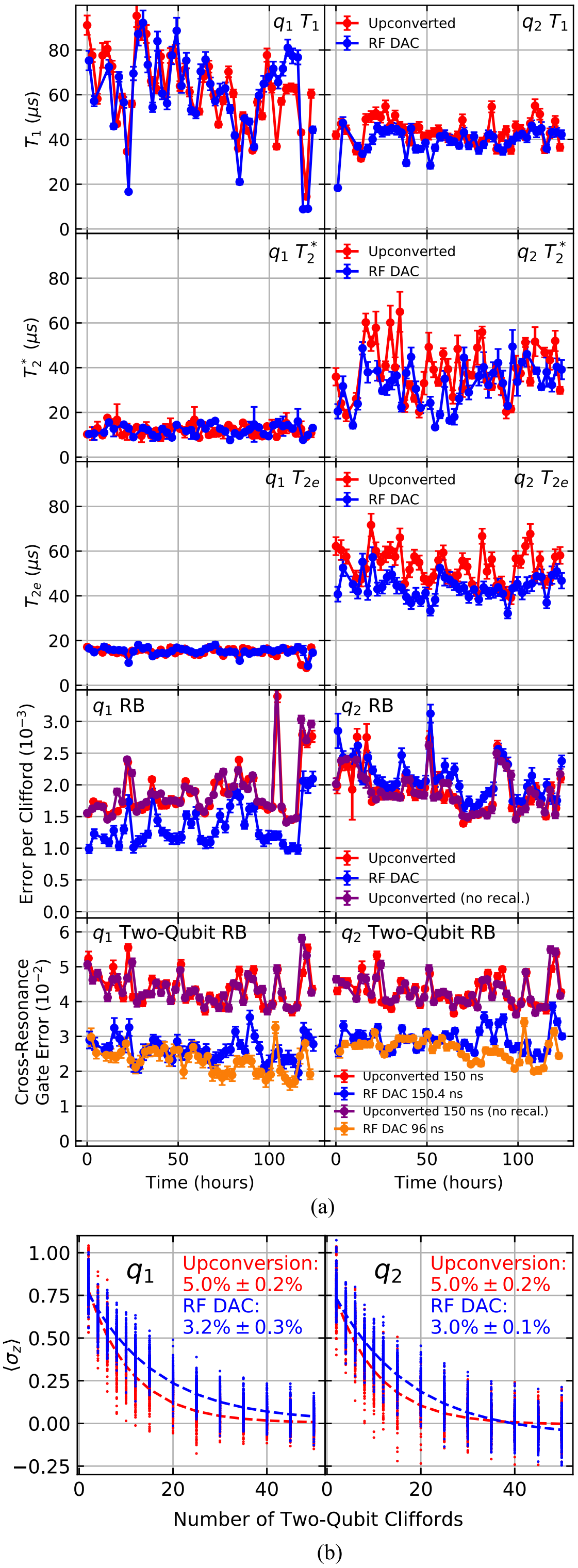

**FIGURE 7. (a) Additional metrics tracked alongside CR gate error over a long timescale. $T_1$ is measured as the characteristic decay time for a qubit in the excited state to relax to the ground state, $T_2^*$ ($T_{2e}$) is measured as the characteristic decay time of a Ramsey (spin echo) experiment, and single-qubit error per Clifford is measured using RB. Significant changes in coherence, single-qubit error, and two-qubit error are consistent between the RF DAC and the upconversion system, indicating that these events are not due to effects of either system. (b) Two-qubit RB data (circles) and exponential fits (dashed lines) corresponding to a random data point in the bottom row of (a).**

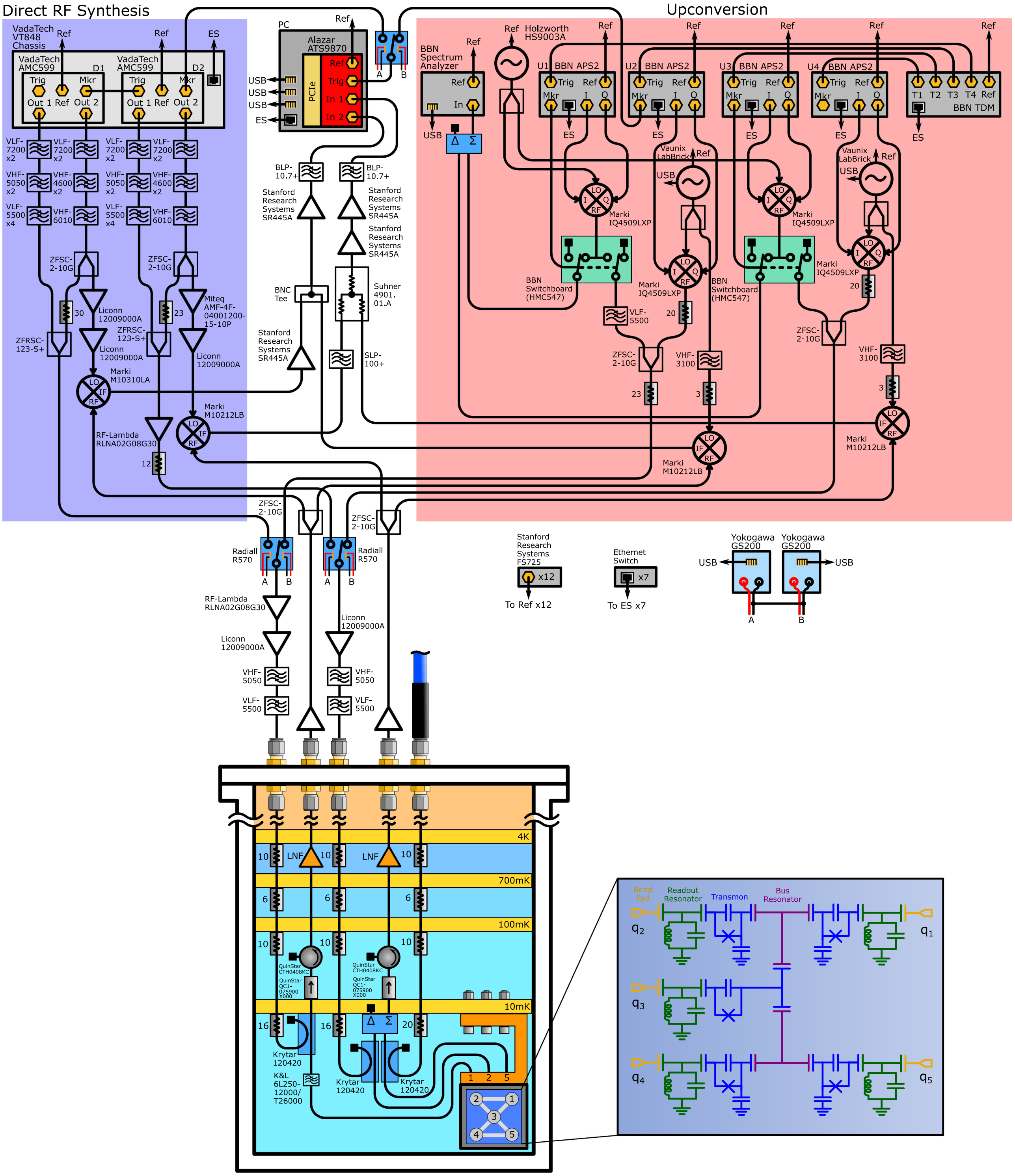

**FIGURE 8.** **Complete experimental setup for two-qubit gate measurements. Microwave switches allow interleaving experiments with each control system to decouple the effect of natural time-varying fluctuations in qubit performance. The VadaTech AMC599 modules integrate two AD9164 RF DACs; Out 1 (2) is used for synthesizing control (readout) pulses. We load the VT848 chassis with two AMC599 modules; D1 interfaces with $q_1$ (the target qubit in the CR gate) and D2 interfaces with $q_2$ (the control qubit). In comparison, separate BBN APS2 units are needed for generating each pulse at baseband, which are synthesized as IQ pairs and upconverted with a Holzworth HS9003A at 5.4 GHz. U1 (U2) synthesizes control (readout) pulses for $q_1$, and U3 (U4) synthesizes control (readout) pulses for $q_2$. We note that there is increased attenuation at the mixing chamber compared to Fig. 9 because the single-qubit and two-qubit data were collected in separate cooldowns approximately six months apart; as such, the coupling to the qubits changed slightly, and we found that additional attenuation was necessary to prevent thermal effects from interfering with gate quality. For components displayed with a part number but no vendor name, the vendor is Mini-Circuits.**

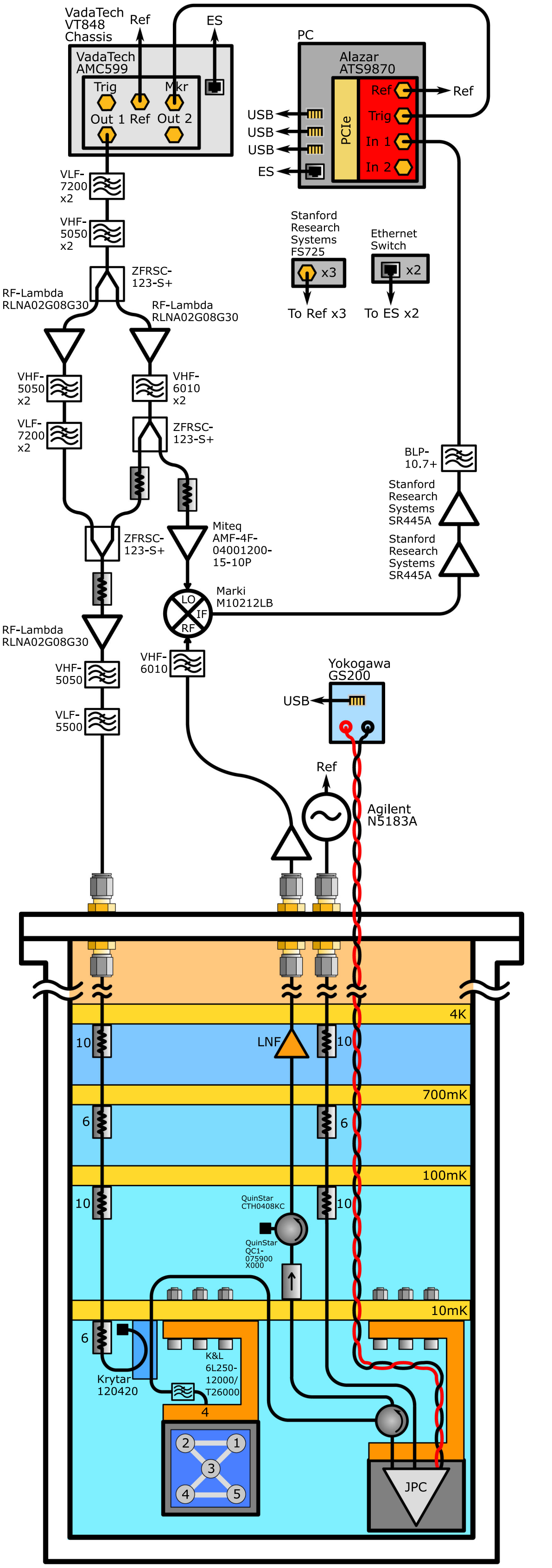

**FIGURE 9.** **Complete experimental setup for single-channel control and readout. For components displayed with a part number but no vendor name, the vendor is Mini-Circuits.**

The monotonic increase in SFDR as a function of tone amplitude displayed by the RF DAC indicates that its output is linear over the range of amplitudes measured, which are representative of realistic amplitudes for CR gates. In contrast, the upconversion system enters a regime where SFDR decreases as amplitude is increased, indicating the presence of nonlinear distortion. Comparing to Fig. 5, the lower SFDR of the RF DAC does not degrade gate performance compared to the upconversion system, and the amplitude at which upconversion SFDR measurably decreases approximately corresponds to the amplitude at which CR gate error begins to increase.

These results highlight the sensitivity of CR gate error to miniscule spurious emissions resulting specifically from nonlinearity, becoming prohibitively large despite the fact that these emissions are suppressed by over 50 dB compared to the synthesized tone. Our observations demonstrate that the distinction between nonlinear harmonics and other types of spurious emissions is critical when considering the electrical characteristics of a potential control system.

We note that when calculating SFDR for the upconversion system, we exclude the suppressed sideband and LO leakage, the relative strengths of which we display separately. This choice stems from our observations that these tones would dominate the SFDR, obfuscating the presence of low-power (yet clearly detrimental) harmonics arising from nonlinearity. Likewise, for the RF DAC, we exclude the tone in the second Nyquist zone as well as weak leakage of the DAC clock. Since it is typical that one must plan carefully to avoid overlap between the suppressed sideband/LO and sensitive frequency locations on the device when using an upconversion system, one must also avoid overlap of qubit frequencies with tones in other Nyquist zones and the DAC clock when using an RF DAC. However, because these emissions may be chosen to be far from the intended tone due to the wide bandwidth of the RF DAC, they are easily filtered out even when the device requires a wide range of tone frequencies. We also note that the decreased SFDR of the RF DAC is primarily due to a lower output amplitude and low-energy, constant-frequency emissions likely resulting from digital switching of the physically nearby FPGA, which may be mitigated by careful engineering of the electronics in a custom solution.

## APPENDIX C
## DEVICE DETAILS AND TWO-QUBIT EXPERIMENTAL CONFIGURATION

We use a device with five fixed-frequency transmons (denoted $q_1$–$q_5$) located at the base of a Bluefors dilution refrigerator (the device is the same used in [7]; two-qubit gate experiments were conducted in a separate cooldown). On the device, one central qubit ($q_3$) is coupled to two bus resonators; one resonator couples $q_3$ to $q_1$ and $q_2$, and the other couples $q_3$ to $q_4$ and $q_5$. Each qubit is dispersively coupled to a readout resonator through which single-qubit gates and two-qubit CR gates are driven. In addition, $q_1$ and $q_4$ have off-chip Purcell filters.

The two-qubit gates under examination use $q_1$ as the target and $q_2$ as the control, whose $|0\rangle \rightarrow |1\rangle$ transition frequencies are 5.3505 and 5.4735 GHz, respectively. They are coupled to readout resonators with frequencies 6.5138 and 6.4616 GHz, respectively. In addition to two-qubit gate error, coherence and single-qubit gate error (as measured by RB) are tracked over the course of the experiment (see Fig. 7). As with two-qubit error, single-qubit error and coherence are consistent between the RF DAC and the upconversion system, indicating that changes in these parameters are not due to deleterious effects of either control system.

Fig. 8 displays the complete experimental setup for all measurements involving two-qubit gates. To decouple the effects of natural time-varying fluctuations in coherence and gate error, experiments using the RF DAC and the upconversion system are interleaved. Two single-pole double-throw microwave switches are used to control the signal source for the two qubits, and a third switch is used to toggle the measurement trigger. We choose mechanical latching microwave switches controlled by DC power supplies, as other types of switches with integrated control electronics were found to degrade qubit coherence. The return line for each qubit is split and distributed to the downconversion circuitry for each control system, and the baseband measurement line for each system is combined and amplified before entering a DC-coupled digitizer. All instruments are locked to a 10-MHz reference generated by a rubidium frequency standard.

We exploit the fact that commercial microwave low-pass filters are not ideal brick-wall low-pass filters, and instead provide 20–30 dB of attenuation in the stopband. Since this is approximately the level of attenuation needed for dispersive readout (based on the power of the output signal), we use low-pass filters with cutoff frequencies in between the control and readout bands as a means of simultaneously attenuating the readout tone without sacrificing dynamic range and protecting against high-frequency amplifier noise. Additional filters at the RF DAC outputs protect against crosstalk (and its subsequent amplification by the gate drive amplifiers) and suppress tones in other Nyquist zones.

The upconversion system is comprised of an array of BBN second-generation Arbitrary Pulse Synthesizers (APS2) [36] and a Holzworth HS9003A ultra-low-phase-noise microwave source. Each APS2 synthesizes an IQ quadrature pair at 1.2 GSa/s, which is then upconverted with a shared LO signal derived from a single channel of the microwave source. A custom switchboard routes the output of the control upconversion mixers to either the qubit drive path or an input on a shared hybrid junction. The $\Sigma$ port of the hybrid junction is connected to a BBN spectrum analyzer; this allows calibration of the mixers to ensure proper suppression of LO leakage and the undesired sideband. For each qubit, one APS2 unit is used for synthesizing control pulses and another is used for synthesizing measurement tones. A BBN Trigger Distribution Module (TDM) ensures synchronized triggering between all APS2 units.

## APPENDIX D
## SINGLE-CHANNEL CONTROL AND READOUT CONFIGURATION

A complete diagram of the setup implementing single-channel control and readout is shown in Fig. 9. In this configuration, we conduct experiments using a single qubit on the device. We choose $q_4$, which has a $|0\rangle \rightarrow |1\rangle$ transition frequency of 5.3622 GHz and is coupled to a readout resonator at 6.5705 GHz. A Josephson parametric converter (JPC) [49] located in a separate shielding can in the mixing chamber is used for enhancing the readout fidelity of the qubit.

As described in the main text, the control signal is split and selectively filtered so that the measurement tone may be recovered for downconversion. The resultant DC signal is then amplified and filtered before entering a digitizer. After splitting the output channel, additional attenuation and amplification are used to maximize output power without entering nonlinear regimes of the amplifiers. The JPC is biased with a DC power supply and pumped with an additional microwave source.

## ACKNOWLEDGMENT

The authors would like to thank M. Gustafsson and G. E. Rowlands for useful discussions, B. Hassick and A. Kreider for experimental assistance, C. A. Ryan and B. Johnson for initial investigation of direct synthesis hardware, and M. Takita, A. D. Córcoles, B. Abdo, and J. M. Chow for work on the superconducting quantum processor. The views and conclusions contained herein are those of the authors and should not be interpreted as necessarily representing the official policies or endorsements, either expressed or implied, of the Office of the Director of National Intelligence, Intelligence Advanced Research Projects Activity, or the U.S. Government. This article does not contain technology or technical data controlled under either the U.S. International Traffic in Arms Regulations or the U.S. Export Administration Regulations.